\documentclass[prd,twocolumn,groupedaddress,showpacs,nofootinbib]{revtex4}
\usepackage{graphicx}
\usepackage{dcolumn}
\usepackage{amssymb}
\usepackage{mathrsfs}
\usepackage{amsmath}
\usepackage{epsfig}
\usepackage[dvips]{color}
\usepackage{hhline}

\begin{document}

\title{ Leptogenesis with testable Dirac neutrino mass generation}

\author{Pei-Hong Gu}

\email{peihong.gu@sjtu.edu.cn}

\affiliation{School of Physics and Astronomy, Shanghai Jiao Tong University, 800 Dongchuan Road, Shanghai 200240, China}

\begin{abstract}

A TeV-scale Higgs doublet can acquire a tiny vacuum expectation value via its small mixing with the standard model Higgs doublet. Ones then can realize a testable Dirac neutrino mass generation through the sizable Yukawa couplings among this new Higgs doublet, several right-handed neutrinos and the standard model lepton doublets. We show the small mixing between the two Higgs doublets can come from certain interactions for generating the cosmic baryon asymmetry.

\end{abstract}


\maketitle

\section{Introduction}

The phenomena of neutrino oscillations have been established by the atmospheric, solar, accelerator and reactor neutrino experiments \cite{pdg2018}. This means three flavors of neutrinos should be massive and mixed. Meanwhile, the cosmological observations have indicated that the neutrinos should be extremely light \cite{pdg2018}. The tiny neutrino masses can be naturally induced at tree level by the so-called type-I \cite{minkowski1977}, type-II \cite{mw1980} and type-III \cite{flhj1989} seesaw extensions of the standard model (SM). Alternatively, the neutrino masses can be achieved in some radiative seesaw models \cite{ma2006}. In these popular seesaw scenarios, the neutrino masses originate from some lepton-number-violating interactions and hence the neutrinos have a Majorana nature. However, we should keep in mind that the theoretical assumption of the lepton number violation and then the Majorana neutrinos has not been confirmed by any experiments. So it is worth studying the possibility of Dirac neutrinos. In analogy to the conventional seesaw mechanisms for the Majorana neutrinos, we can construct the type-I \cite{rw1983}, type-II \cite{gh2006} and type-III \cite{gu2016} Dirac seesaw as well as the radiative Dirac seesaw \cite{gs2007} for the Dirac neutrinos. In the Majorana or Dirac seesaw models, the cosmic baryon asymmetry, which is another big challenge to the SM, can be understood in a natural way. This is the so-called leptogenesis mechanism \cite{fy1986} and has been widely studied \cite{lpy1986,fps1995,ms1998,bcst1999,hambye2001,di2002,gnrrs2003,hs2004,bbp2005,ma2006,dlrw1999,mp2002,tt2006,gh2006,gu2016,gs2007}

In this paper we shall present a double Dirac seesaw scenario to simultaneously generate the tiny neutrino masses and the cosmic baryon asymmetry. Besides the SM gauge symmetries, our model respects an additionally global or gauge symmetry, under which the right-handed neutrinos have no Yukawa couplings with the SM. After spontaneously breaking this additional symmetry, two or more heavy Higgs singlets can acquire their suppressed vacuum expectation values (VEVs) and then a second Higgs doublet, which realizes the Yukawa couplings of the right-handed neutrinos to the SM lepton doublets, can obtain a small mixing with the SM Higgs doublet. Therefore, the second Higgs doublet can also acquire a suppressed VEV even if it is set at the TeV scale. This means a testable Dirac neutrino mass generation \cite{wwy2006,dl2010}. Furthermore, the heavy Higgs singlet decays can produce an asymmetry stored in the second Higgs doublet. This asymmetry can result in a desired baryon asymmetry in association with the sphaleron processes \cite{krs1985}.

\section{The models}

We denote the SM fermions and Higgs by
\begin{eqnarray}
\begin{array}{c}q_{L}^{}(3,2,+\frac{1}{6})\end{array}\!\!\!&=&\!\!\left[\begin{array}{c} u^{}_{L} \\
[1mm] d_{L}^{}\end{array}\right]\,,~~\begin{array}{c}d_R^{}(3,1,-\frac{1}{3})\,,\end{array}~~\begin{array}{c}u_R^{}(3,1,+\frac{2}{3})\,,\end{array}\nonumber\\
[1mm]
\begin{array}{c}l_{L}^{}(1,2,-\frac{1}{2})\end{array}\!\!\!&=&\!\!\left[\begin{array}{c} \nu^{}_{L} \\
[1mm] e_{L}^{}\end{array}\right]\,,~~\begin{array}{c}e_{R}^{}(1,1,-1)\,,\end{array}\nonumber\\
[1mm]
\begin{array}{c}\phi(1,2,-\frac{1}{2})\end{array}\!\!\!&=&\!\!\left[\begin{array}{c} \phi^{0}_{} \\
[1mm] \phi^{-}_{}\end{array}\right]\,.
\end{eqnarray}
Here and thereafter the brackets following the fields describe the transformations under the $SU(3)_c^{}\times SU(2)_L^{}\times U(1)^{}_{Y}$ gauge groups. The SM charged fermions can obtain their masses through the Yukawa interactions as follows, 
\begin{eqnarray}
\label{yukawa1}
\mathcal{L} &\supset&  - y_d^{} \bar{q}_L^{} \tilde{\phi} d_R^{} - y_u^{} \bar{q}_L^{} \phi u_R^{} - y_e^{} \bar{l}_L^{} \tilde{\phi} e_R^{} +\textrm{H.c.}\,.
\end{eqnarray}

Similarly, we can introduce two or more right-handed neutrinos,
\begin{eqnarray}
\nu_R^{}(1,1,0)\,,
\end{eqnarray}
to construct the Yukawa interactions for generating a neutrino mass matrix with at least two nonzero eigenvalues,
\begin{eqnarray}
\label{yukawa2}
\mathcal{L} &\supset&  -  y_\nu^{} \bar{l}_{L}^{} \phi \nu_{R} +\textrm{H.c.}\,.
\end{eqnarray}
In this Dirac neutrino scenario, the tiny neutrino masses would enforce the Yukawa couplings $y_\nu^{}$ to be extremely small. This smallness could be naturally explained by certain Dirac seesaw mechanisms. In the Dirac seesaw models, the Yukawa interactions (\ref{yukawa2}) will not appear before a new symmetry is spontaneously broken.

The present work will be based on the models as below,
\begin{eqnarray}
\label{lag}
\mathcal{L} \!\!&\supset&\!\! - M_{a}^{2}\sigma_a^\dagger \sigma_a^{} - (\mu_\eta^2 + \lambda_{\xi\eta}^{}\xi^\dagger_{}\xi + \lambda_{\phi\eta}^{} \phi^\dagger_{} \phi +\lambda_{a \eta}^{} \sigma_a^\dagger \sigma_a^{}) \eta^\dagger_{}\eta 
\nonumber\\
\!\!&&\!\!-\lambda'^{}_{\phi\eta}\eta^\dagger_{}\phi\phi^\dagger_{}\eta- \kappa_a^{} \sigma_a^{} \xi^\dagger_{} \xi^\dagger_{} \xi^\dagger_{} -  \rho_a^{} \sigma_a^{} \eta^\dagger_{}\phi -  f\bar{l}_{L}^{} \eta \nu_{R} \nonumber\\
\!\!&&\!\!+\textrm{H.c.}\,,~~(a=1,...,n\geq 2)\,,
\end{eqnarray}
where $\sigma$ and $\xi$ are the SM singlets while $\eta$ is a new Higgs doublet, 
\begin{eqnarray}
\begin{array}{c}\sigma(1,1,0)\,,\end{array}~~\begin{array}{c}\xi(1,1,0)\,,\end{array}~~\begin{array}{c}\eta(1,2,-\frac{1}{2})\end{array}\!\!=\!\left[\begin{array}{c} \eta^{0}_{} \\
[1mm] \eta^{-}_{}\end{array}\right]\,.
\end{eqnarray}
Note that one of the four parameters $\kappa_{a,b\neq a}^{}$ and $\rho_{a,b\neq a}^{}$ can always keep to be complex.

In order to forbid the Yukawa interactions (\ref{yukawa2}) and then construct the model (\ref{lag}), we can introduce a $U(1)_{B-L}^{}$ gauge symmetry under which three right-handed neutrinos $\nu_{R1}^{}$, $\nu_{R2}^{}$ and $\nu_{R3}^{}$ carry the $B-L$ numbers $-4$, $-4$ and $+5$ \cite{mp2007}, respectively, while the new scalars $\eta$, $\xi$ and $\sigma$ carry the $B-L$ numbers $+3$, $+1$ and $+3$, respectively. In this $U(1)_{B-L}^{}$ scenario, although the third right-handed neutrino $\nu_{R3}^{}$ has no Yukawa couplings and hence keeps massless, it can decouple at a temperature above the QCD scale and hence can escape from the BBN constraint when the $U(1)_{B-L}^{}$ symmetry is broken above the TeV scale. Alternatively, we can consider a $U(1)_X^{}$ global symmetry under which only the non-SM fields are nontrivial, i.e.
\begin{eqnarray}
\label{xnumber}
(+3,-3,-1,-3)\quad \textrm{for}\quad (\nu_R^{},\eta,\xi,\sigma)\,.
\end{eqnarray}
Remarkably, this $U(1)_X^{}$ global symmetry can be always allowed even if we have introduced the previous $U(1)_{B-L}^{}$ gauge symmetry. 

It is easy to see that in Lagrangian (\ref{lag}), we can introduce an arbitrary definition of the global lepton numbers of the right-handed neutrinos $\nu_R^{}$ and the new scalars $(\eta,\xi,\sigma)$ to conserve the global lepton number, i.e.
\begin{eqnarray}
\begin{array}{c}\nu_R^{}(x)\,,~~\eta(1-x)\,,~~\xi(\frac{1-x}{3})\,,~~\sigma(1-x)\,.\end{array}
\end{eqnarray}
We will show later our leptogenesis mechanism does not depend on the parameter $x$.

\section{Dirac neutrino mass}

The Higgs singlet $\xi$ is responsible for spontaneously breaking the $U(1)_{B-L}^{}$ gauge symmetry or the $U(1)_X^{}$ global symmetry. After this symmetry breaking, the heavy Higgs singlets $\sigma$ can acquire their suppressed VEVs,
\begin{eqnarray}
\langle \sigma_a^{} \rangle \simeq -\frac{\kappa_a^{}  \langle \xi\rangle^3_{}}{M_a^2} \ll \langle \xi \rangle ~~\textrm{for}~~M_a^2\gg \langle\xi\rangle^2_{}\,. 
\end{eqnarray}
As a result, the second Higgs doublet $\eta$ can have a small mixing with the SM Higgs doublet $\phi$, i.e.
\begin{eqnarray}
\mathcal{L}\supset - \mu^2_{\eta\phi} \eta^\dagger_{}\phi +\textrm{H.c.}~~\textrm{with}~~\mu_{\eta\phi}^2 = \rho_a^{}\langle\sigma_a^{}\rangle\,.
\end{eqnarray}
This means the second Higgs doublet $\eta$ can also pick up a suppressed VEV after the SM Higgs doublet $\phi$ drives the electroweak symmetry breaking, i.e.
\begin{eqnarray}
\langle \eta \rangle \simeq -\frac{\rho_a^{}  \langle \sigma_a^{} \rangle \langle \phi\rangle }{m_{\eta^0_{}}^2} \ll \langle \phi \rangle\,. 
\end{eqnarray}
Here $m_{\eta^0_{}}^{}$ is the mass of the neutral component $\eta^0_{}$ of the second Higgs doublet $\eta$, i.e.
\begin{eqnarray}
m_{\eta^0_{}}^2 = \mu_\eta^2 + \lambda_{\xi\eta}^{}\langle\xi \rangle^2_{}+ (\lambda_{\phi\eta}^{}+\lambda'^{}_{\phi\eta}) \langle \phi \rangle^2_{}+\lambda_{a \eta}^{} \langle \sigma_a^{}\rangle^2_{}\,,
\end{eqnarray}
which could have a split with the mass $m_{\eta^\pm_{}}^{}$ of the charged component $\eta^{\pm}_{}$, i.e.
\begin{eqnarray}
m_{\eta^\pm_{}}^2 = \mu_\eta^2 + \lambda_{\xi\eta}^{}\langle\xi \rangle^2_{}+ \lambda_{\phi\eta}^{}\langle \phi \rangle^2_{}+\lambda_{a \eta}^{} \langle \sigma_a^{}\rangle^2_{}\,.
\end{eqnarray}

\begin{figure}
\vspace{8cm} \epsfig{file=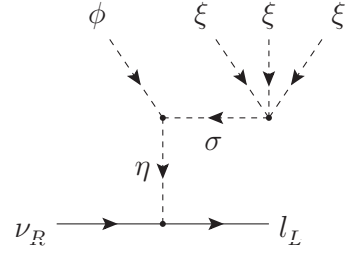, bbllx=5cm, bblly=6.0cm,
bburx=15cm, bbury=16cm, width=8cm, height=8cm, angle=0,
clip=0} \vspace{-11.5cm} \caption{\label{numass} The Dirac neutrino mass generation.}
\end{figure}

Through the Yukawa interactions of the second Higgs doublet $\eta$ to the right-handed neutrinos $\nu_R^{}$ and the SM lepton doublets $l_L^{}$, we now can obtain a tiny neutrino mass term,
\begin{eqnarray}
\mathcal{L} &\supset& -m_{\alpha i}^{} \bar{\nu}_{L\alpha}^{} \nu_{Ri}^{} + \textrm{H.c.} ~~\textrm{with}~~m_{\alpha i}^{} = f_{\alpha i}^{} \langle \eta\rangle\,.
\end{eqnarray}
This Dirac neutrino mass generation depends on two-step suppression of the VEVs $\langle\sigma\rangle$ and $\langle\eta\rangle$ so that it may be titled as a double type-II Dirac seesaw mechanism, in analogy to our double type-II seesaw \cite{ghsz2009} for the Majorana neutrinos. We can conveniently understand this double Dirac seesaw in Fig. \ref{numass}.

\section{Heavy Higgs singlet decays}

\begin{figure*}
\vspace{7.5cm} \epsfig{file=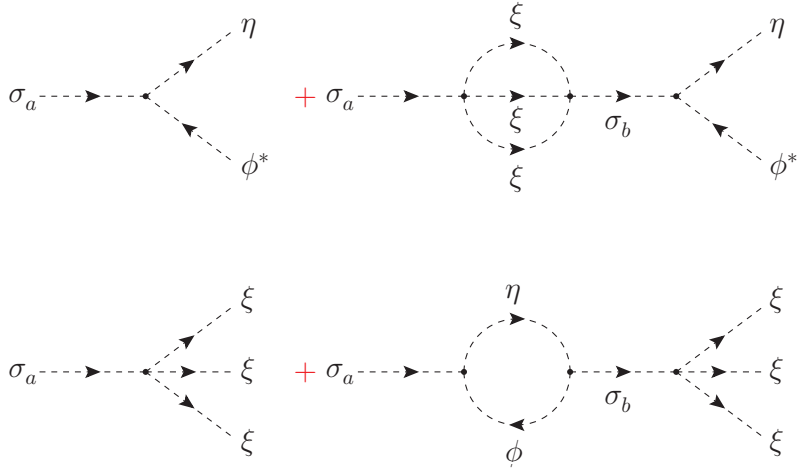, bbllx=5cm, bblly=6.0cm,
bburx=15cm, bbury=16cm, width=8cm, height=8cm, angle=0,
clip=0} \vspace{-8.5cm} \caption{\label{sdecay} The heavy Higgs singlet decayss.}
\end{figure*}

As shown in Fig. \ref{sdecay}, the heavy Higgs singlets $\sigma_a^{}$ have two decay modes,
\begin{eqnarray}
\sigma_a^{}\rightarrow \xi\xi\xi\,,~~ \sigma_a^{} \rightarrow \eta \phi^\ast_{}\,.
\end{eqnarray}
As long as the CP is not conserved, we can expect a CP asymmetry in the above decays, 
\begin{eqnarray}
\varepsilon_a^{} &=& \frac{ \Gamma(\sigma_a^{} \rightarrow \eta \phi^\ast_{}) - \Gamma(\sigma_a^{\ast} \rightarrow \eta^\ast_{} \phi ) }{\Gamma_a^{}}\nonumber\\
&=&- \frac{\Gamma(\sigma_a^{}\rightarrow \xi\xi\xi) - \Gamma(\sigma_a^{\ast}\rightarrow \xi^\ast_{}\xi^\ast_{}\xi^\ast_{}) }{\Gamma_a^{}}\neq 0\,,\end{eqnarray}
where $\Gamma_a^{} $ is the total decay width,
\begin{eqnarray}
\Gamma_a^{} &=& \Gamma(\sigma_a^{}\rightarrow \xi\xi\xi) + \Gamma(\sigma_a^{} \rightarrow \eta \phi^\ast_{}) \nonumber\\
&=&  \Gamma(\sigma_a^{\ast}\rightarrow \xi^\ast_{}\xi^\ast_{}\xi^\ast_{}) + \Gamma(\sigma_a^{\ast} \rightarrow \eta^\ast_{} \phi ) \,.
\end{eqnarray}
We can calculate the decay width at tree level and the CP asymmetry at one-loop level, 
\begin{eqnarray}
\Gamma_a^{} &=&\frac{1}{8\pi}\left(\frac{|\rho_a^{}|^2}{M_{a}^2}+\frac{3|\kappa_a^{}|^2_{}}{32\pi^2_{}}\right) M_{a}^{}\,,\\
\varepsilon_a^{} &=&- \frac{3}{64\pi^3_{}}\sum_{b\neq a}^{}\frac{\textrm{Im}\left(\kappa^\ast_{a}\kappa_b^{}\rho_a^{}\rho_b^\ast\right)}{\frac{|\rho_a^{}|^2}{M_{a}^2}+\frac{3|\kappa_a^{}|^2_{}}{32\pi^2_{}}}\frac{1}{M_b^2-M_a^2}\nonumber\\
&=&- \frac{3}{64\pi^3_{}}\sum_{b\neq a}^{}\frac{|\kappa^{}_{a}\kappa_b^{}\rho_a^{}\rho_b^{}|\sin\delta_{ab}^{}}{\frac{|\rho_a^{}|^2}{M_{a}^2}+\frac{3|\kappa_a^{}|^2_{}}{32\pi^2_{}}}\frac{1}{M_b^2-M_a^2}\,.
\end{eqnarray}
Here $\delta_{ab}^{}$ is the relative phase among the parameters $\rho_{a,b}^{}$ and $\kappa_{a,b}^{}$.

After the heavy Higgs singlets $\sigma_a^{}$ go out of equilibrium, their decays can generate an $X$ asymmetry $X_\eta^{}$ and a lepton asymmetry $L_\eta^{}$ stored in the second Higgs doublet $\eta$. For demonstration, we simply assume a hierarchical spectrum of the heavy Higgs singlets $\sigma_a^{}$, i.e. $M_{\sigma_1^{}}^{2}\ll M_{\sigma_{2,...}^{}}^2$. In this case, the decays of the lightest $\sigma_1^{}$ should dominate the final asymmetries in the second Higgs doublet $\eta$, i.e.
\begin{eqnarray}
X_\eta^{}&=& -3 \varepsilon_1^{}\left(\frac{n^{eq}_{\sigma_1} }{s}\right)\left|_{T=T_D^{}}^{}\right.\,,\nonumber\\
L_\eta^{}&=& (1-x)  \varepsilon_1^{}\left(\frac{n^{eq}_{\sigma_1} }{s}\right)\left|_{T=T_D^{}}^{}\right.\,,
\end{eqnarray}
where the factors $-3$ and $1-x$ respectively are the $X$ number and the lepton number of the second Higgs doublet $\eta$, the symbols $n^{eq}_{\sigma_1}$ and $T_D^{}$ respectively are the equilibrium number density and the decoupled temperature of the heavy Higgs singlets $\sigma_1^{}$, while the character $s$ is the entropy density of the universe \cite{kt1990}.

\section{Baryon asymmetry}

The $X$ asymmetry $X_\eta^{}$ and the lepton asymmetry $L_\eta^{}$ stored in the second Higgs doublet $\eta$ will lead to a lepton asymmetry stored in the SM lepton doublets $l_L^{}$ because of the related Yukawa interactions in Eq. (\ref{lag}). The sphaleron processes then can partially transfer this SM lepton asymmetry to a baryon asymmetry.

We now analysize the chemical potentials \cite{ht1990} to discuss the details of these conversions. For this purpose, we denote $\mu_{q}^{}$, $\mu_d^{}$, $\mu_u^{}$, $\mu_l^{}$, $\mu_e^{}$, $\mu_\nu^{}$, $\mu_\phi^{}$ and $\mu_\eta^{}$ for the chemical potentials of the fields $q_{L}^{}$, $d_{R}^{}$, $u_{R}^{}$, $l_L^{}$, $e_R^{}$, $\nu_R^{}$, $\phi$ and $\eta$. We then can consider the chemical potentials in two phases,
\begin{itemize}
\item phase-I: during the inert Higgs singlet decays and the second Higgs doublet decays, 
\item phase-II: during the second Higgs doublet decays and the electroweak symmetry breaking.
\end{itemize}

In phase-I, the SM Yukawa interactions are in equilibrium and hence yield,
\begin{eqnarray}
\label{chemical1}
-\mu_{q}^{}+\mu_{d}^{}-\mu_{\phi}^{}&=&0\,,\\
 -\mu_{q}^{}+\mu_{u}^{}+\mu_{\phi}^{}&=&0\,,\\
 -\mu_{l}^{}+\mu_{e}^{}-\mu_{\phi}^{}&=&0\,,
\end{eqnarray}
the fast sphalerons constrain,
\begin{eqnarray}
\label{chemical2}
3\mu_{q}^{}+\mu_{l}^{}&=&0\,,
\end{eqnarray}
while the neutral hypercharge in the universe requires,
\begin{eqnarray}
\label{chemical3}
3\left( \mu_{q}^{} -\mu_{d}^{}+2\mu_{u}^{}-\mu_{l}^{} -\mu_{e}^{}\right)-2\mu_{\phi}^{}- 2\mu_{\eta}^{} =0\,.
\end{eqnarray}
In addition, the Yukawa interactions involving the right-handed neutrinos are also in equilibrium. This means
\begin{eqnarray}
\label{chemical4}
-\mu_{l}^{}+\mu_{\nu}^{}+\mu_{\eta}^{}=0\,.
\end{eqnarray}
In the above Eqs. (\ref{chemical1}-\ref{chemical4}), we have identified the chemical potentials of the different-generation fermions because the Yukawa interactions establish an equilibrium between the different generations. By solving Eqs. (\ref{chemical1}-\ref{chemical4}), we express the chemical potentials in phase-I as below,
\begin{eqnarray}
&&\mu_\phi^{I}= -\frac{4}{7}\mu_l^{I} - \frac{1}{7}\mu_\eta^{I}\,,~~\mu_q^{I}= -\frac{1}{3} \mu_l^{I}\,, \nonumber\\
&&\mu_d^{I}= -\frac{19}{21}\mu_l^{I}-\frac{1}{7}\mu_\eta^{I} \,,~~\mu_u^{I}= \frac{5}{21}\mu_l^{I}+\frac{1}{7}\mu_\eta^{I} \,,\nonumber\\
&&\mu_e^{I} = \frac{3}{7}\mu_l^{I} -\frac{1}{7}\mu_\eta^{I}\,,~~\mu_{\nu}^{I}=\mu_l^{I} -\mu_\eta^{I}\,.
\end{eqnarray}

Now the baryon number can be given by
\begin{eqnarray}
B^I_{}= 3(2\mu_q^{I} + \mu_d^{I} + \mu_u^{I}) = -4 \mu_l^{I}\,.
\end{eqnarray}
As for the lepton number, it should be 
\begin{eqnarray}
L^{I}_{}= L_{SM}^{I} + L_{\nu_R^{}}^{I}+L_{\eta}^{I}\,,
\end{eqnarray}
with $L_{SM}^{I}$, $L_{\nu_R}^{I}$ and $L_\eta^I$ being the lepton number in the SM leptons, the right-handed neutrinos and the second Higgs doublet, respectively, 
\begin{eqnarray}
L_{SM}^{I}&=& 3(2\mu_l^{I} + \mu_e^{I} ) = \frac{51}{7}\mu_l^{I} - \frac{3}{7}\mu_\eta^{I}\,,\nonumber\\
L_{\nu_R^{}}^{I}&=& n  x\mu_\nu^{I} = nx (\mu_l^{I}-\mu_\eta^{I}) \,,~~(n\geq 2)\,,\nonumber\\
L_\eta^{I}&=&4(1-x)\mu_\eta^{I}\,.
\end{eqnarray}
The $B-L$ number then can be computed by  
\begin{eqnarray}
(B-L)^{I}_{}&=& B^{I}_{}- (L_{SM}^{I} + L_{\nu_R^{}}^{I}+L_\eta^{I}) \nonumber\\
&=&-\frac{79+7nx}{7}\mu_l^{I} + \frac{7(n+4)x-25}{7}\mu_\eta^{I}\,.
\end{eqnarray}
According to the $U(1)_X^{}$ global symmetry (\ref{xnumber}), we can also obtain an $X$ number, 
\begin{eqnarray}
X^I_{}&=&X_{\nu_R^{}}^{I}+X_\eta^{I}= 3n \mu_{\nu}^{I} - 12 \mu_\eta^{I}\nonumber\\
&=&3n\mu_l^I - 3(n+4)\mu_\eta^I\,.
\end{eqnarray}

After the heavy Higgs singlet decays, the $B-L$ and $X$ numbers in the SM fields, the right-handed neutrinos and the second Higgs doublet should be both conserved. Therefore, we can read
\begin{eqnarray}
(B-L)^{I}_{}&=& (B-L)^{i}_{}\nonumber\\
&=& - L^{i}_\eta =-(1-x) \varepsilon_1^{}\left(\frac{n^{eq}_{\sigma_1} }{s}\right)\left|_{T=T_D^{}}^{}\right. \,, \nonumber\\
X^I_{}&=& X^i_\eta = -3 \varepsilon_1^{}\left(\frac{n^{eq}_{\sigma_1} }{s}\right)\left|_{T=T_D^{}}^{}\right.\,, 
\end{eqnarray}
where $L^i_\eta$ and $X^i_\eta$ are the initial $B-L$ and $X$ numbers from the decays of the heavy Higgs singlets into the second Higgs doublet. Clearly, we should have
\begin{eqnarray}
L^{i}_\eta = -\frac{1-x}{3} X^i_\eta \,. 
\end{eqnarray}
So we eventually can derive
\begin{eqnarray}
B^{I}_{}&=&\frac{7n+3}{3(26n+79)}X^{i}_{\eta}\,,\nonumber\\
L^{I}_{SM}&=&-\frac{4n-1}{26n+79}X^{i}_{\eta}\,,\nonumber\\
L_{\nu_R^{}}^{I}&=&\frac{19nx}{3(26n+79)}X^{i}_{\eta}\,,\nonumber\\
L_\eta^{I}&=&-\frac{(7n+79)(1-x)}{3(26n+79)}X^{i}_{\eta}\,.
\end{eqnarray}

In phase-II, the second Higgs doublet $\eta$ has already decayed so that the condition (\ref{chemical3}) for the zero hypercharge should be modified by, 
\begin{eqnarray}
\label{chemical7}
3\left(\mu_{q}^{}-\mu_{d}^{}+2\mu_{u}^{}- \mu_{l}^{}-\mu_{e}^{}\right)-2\mu_{\phi}^{}=0\,.
\end{eqnarray}
Ones then can solve Eqs. (\ref{chemical1}-\ref{chemical2}) and (\ref{chemical7}) to determine the chemical potentials,
\begin{eqnarray}
\label{chemical8}
&&\mu_q^{II}=-\frac{1}{3}\mu_l^{II}\,,~~\mu_d^{II}=-\frac{19}{21}\mu_l^{II}\,,~~\mu_u^{II}=\frac{5}{21}\mu_l^{II}\,,\nonumber\\
&&\mu_{e}^{II}=\frac{3}{7}\mu_l^{II}\,,~~\mu_\phi^{II}=-\frac{4}{7}\mu_l^{II}\,.
\end{eqnarray}
At this stage, the baryon and lepton numbers in the SM should be 
\begin{eqnarray}
B^{II}_{}&=&\frac{28}{79}(B-L)^{II}_{}\,,\nonumber\\
L_{SM}^{II}&=&-\frac{51}{79}(B-L)^{II}_{}\,.
\end{eqnarray}
Because the second Higgs doublet carries the lepton number $1-x$, its decays can produce the lepton doublets with a number $L_\eta^I/(1-x)$. The conserved $(B-L)^{II}_{}$ number thus should be
\begin{eqnarray}
(B-L)^{II}_{}=B^{I}_{}-L^I_{SM}-\frac{1}{1-x}L^I_\eta = \frac{1}{3}X^i_\eta\,.
\end{eqnarray}
This means the final baryon asymmetry can be given by 
\begin{eqnarray}
\label{bauf}
B^f_{}=B^{II}_{}=\frac{28}{237} X_\eta^i = -\frac{28}{79} \varepsilon_1^{}\left(\frac{n^{eq}_{\sigma_1} }{s}\right)\left|_{T=T_D^{}}^{}\right.\,,
\end{eqnarray}
which is independent on the choice of the global lepton numbers of the right-handed neutrinos and the other non-SM fields.

\section{Numerical example}

For demonstration, we consider the weak washout case \cite{kt1990}, 
\begin{eqnarray}
\label{weak}
\left.\left[\Gamma_{1}^{}< H(T)=\left(\frac{8\pi^{3}_{}g_{\ast}^{}}{90}\right)^{\frac{1}{2}}_{}\frac{T^2_{}}{M_{\textrm{Pl}}^{}}\right]\right|_{T=M_1^{}}^{}\,.
\end{eqnarray}
Here $H(T)$ is the Hubble constant with $M_{\textrm{Pl}}^{}\simeq 1.22\times 10^{19}_{}\,\textrm{GeV}$ being the Planck mass and $g_{\ast}^{}=112.75+1.75n$ being the relativistic degrees of freedom (the SM fields plus the right-handed neutrinos $\nu_{R}^{}$, the second Higgs doublet $\eta$ and the Higgs singlet $\xi$.). The final baryon asymmetry (\ref{bauf}) then can approximate to  
\begin{eqnarray}
B^f_{}\sim -\frac{28}{79} \frac{\varepsilon_1^{}}{g_\ast^{}}\,.
\end{eqnarray}

As an example, we assume three right-handed neutrinos and then obtain
\begin{eqnarray}
\frac{\Gamma_1^{}}{H(T)}\left|_{T=M_1^{}}^{}\right.&\simeq &0.45\,,\nonumber\\
\varepsilon_1^{}&=&-7.8\times 10^{-6} \sin\delta_{12}^{}\,,\nonumber\\
B^f_{}&\sim& 10^{-10}_{} \left(\frac{\sin\delta_{12}^{}}{0.005}\right)\,,
\end{eqnarray}
by inputting
\begin{eqnarray}
&&M_{1}^{}=10^{14}_{}\,\textrm{GeV}\,,~|\rho_1^{}|=3\times 10^{12}_{}\,\textrm{GeV}\,, ~|\kappa_1^{}|=0.3\,;\nonumber\\
&&M_{2}^{}=10^{15}_{}\,\textrm{GeV}\,,~|\rho_2^{}| =3\times 10^{13}_{}\,\textrm{GeV}\,,~|\kappa_2^{}|=0.3\,.\nonumber\\
&&
\end{eqnarray}
We further take  
\begin{eqnarray}
\langle\xi\rangle= \mathcal{O}(1-10\,\textrm{TeV})\,,~~m_{\eta^0}^{}=\mathcal{O}(\textrm{TeV})\,,
\end{eqnarray}
to realize
\begin{eqnarray}
\langle \eta \rangle=\mathcal{O}(0.01-10\,\textrm{eV})\,.
\end{eqnarray}
and hence
\begin{eqnarray}
m_\nu^{}= \mathcal{O}(\textrm{0.01-0.1\,eV})~~\textrm{for}~~f=\mathcal{O}(0.001-1)\,.
\end{eqnarray}
So, the above parameter choice can explain the baryon asymmetry and the neutrino masses.

\section{Conclusion}

In this paper we have demonstrated a double Dirac seesaw mechanism. In our scenario, the heavy Higgs singlets can suppress the mixing between the second Higgs doublet and the SM Higgs doublet after the additionally gauge or global symmetry is spontaneously broken. Therefore, the second Higgs doublet can naturally pick up a tiny VEV even if it is set at the TeV scale. Through the sizable Yukawa couplings of the SM lepton doublets to the right-handed neutrinos and the second Higgs doublet, we can realize a testable Dirac neutrino mass generation. Furthermore, the interactions for the small mixing between the two Higgs doublets can also explain the observed baryon asymmetry in association with the sphaleron processes.

\textbf{Acknowledgement}: This work was supported by the National Natural Science Foundation of China under Grant No. 11675100 and the Recruitment Program for Young Professionals under Grant No. 15Z127060004.

\end{document}